\documentclass[aps,prl,twocolumn,superscriptaddress,longbibliography]{revtex4-1}
\usepackage[utf8]{inputenc}
\usepackage{graphicx}
\DeclareGraphicsExtensions{.png,.jpg,.eps,.pdf}

\usepackage{float}
\usepackage{xcolor}
\usepackage{amsmath}
\usepackage{float}
\usepackage[colorlinks=true,citecolor=blue]{hyperref}

\DeclareMathAlphabet\mathbfcal{OMS}{cmsy}{b}{n}

\begin{document}

\title{Self-doped flat band and spin-triplet superconductivity in monolayer 1T-TaSe$_{2-x}$Te$_{x}$}

\author{Jan Phillips}
  \email{j.phillips@usc.es}
\affiliation{Departamento de F\'{i}sica Aplicada,
  Universidade de Santiago de Compostela, E-15782 Campus Sur s/n,
  Santiago de Compostela, Spain}
\affiliation{Instituto de Materiais iMATUS, Universidade de Santiago de Compostela, E-15782 Campus Sur s/n, Santiago de Compostela, Spain}  

\author{Jose L. Lado}
\affiliation{Department of Applied Physics, Aalto University, 02150 Espoo, Finland}

\author{Víctor Pardo}
\affiliation{Departamento de F\'{i}sica Aplicada, Universidade de Santiago de Compostela, E-15782 Campus Sur s/n, Santiago de Compostela, Spain}
\affiliation{Instituto de Materiais iMATUS, Universidade de Santiago de Compostela, E-15782 Campus Sur s/n, Santiago de Compostela, Spain}  
  
\author{Adolfo O. Fumega}
\email{adolfo.oterofumega@aalto.fi}
\affiliation{Department of Applied Physics, Aalto University, 02150 Espoo, Finland}

\begin{abstract}
Two-dimensional van der Waals materials have become an established platform to engineer flat bands which can lead to strongly-correlated emergent phenomena. In particular, the family of Ta dichalcogenides in the 1\textit{T} phase presents a star-of-David charge density wave that creates a flat band at the Fermi level. For TaS$_2$ and TaSe$_2$ this flat band is at half filling leading to a magnetic insulating phase. In this work, we theoretically demonstrate that ligand substitution in the TaSe$_{2-x}$Te$_x$ system produces a transition from the magnetic insulator to a non-magnetic metal in which the flat band gets doped away from half-filling.  
For $x\in[{0.846},{1.231}]$ the spin-polarized flat band is self-doped and the system becomes a magnetic metal. In this regime, we show that attractive interactions promote three different spin-triplet superconducting phases as a function of $x$, corresponding to a nodal f-wave and two topologically-different chiral p-wave superconducting phases. 
Our results establish monolayer TaSe$_{2-x}$Te$_{x}$ as a promising platform for correlated flat band physics leading to unconventional superconducting states.
\end{abstract}

\maketitle

\section{Introduction}
The presence of flat bands around the Fermi level is one of the key ingredients to stabilize exotic orders in quantum materials. In the flat band regime, the kinetic energy of electrons is much smaller than the Coulomb interactions. As a result, the physical behavior of the system is dominated by electronic interactions which lead to strongly correlated phenomena\cite{Mott_1949,PhysRev.115.2,PhysRevLett.22.295,10.1143/PTP.32.37,Peotta2015}.
Traditionally, research has been focused on studying transition metals and rare earth compounds in which flat bands occur naturally due to the localized nature of their d or f electrons\cite{SMITH198383}. 
Modifying the stoichiometry of the parent compound by ligand substitution has been a fruitful strategy to tune the electronic filling and explore complex phase diagrams in systems with localized electrons\cite{Bednorz1986,Li2019,RevModPhys.79.1015,PhysRevLett.121.087203,doi:10.1126/sciadv.abb9379}.

\begin{figure*}
\centering
\includegraphics[width=\textwidth, draft=false]{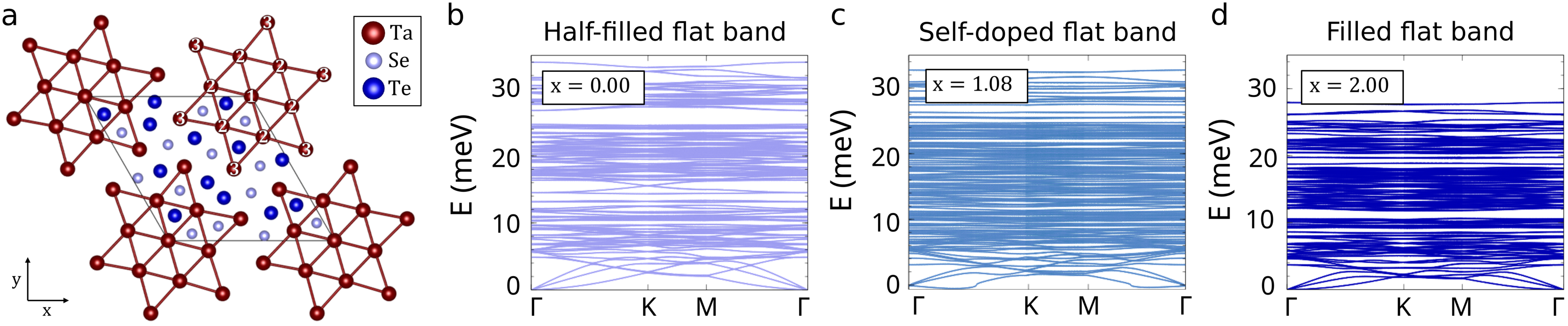}
\caption{(a) Top view of the CDW phase structure of TaSe$_{2-x}$Te$_x$. The numbered Ta atoms refer to the three inequivalent metal atoms in the CDW. Phonon band structures for TaSe$_2$ (b), TaSe$_{2-x}$Te$_x$ for x=1.08 (c) and TaTe$_2$ (d) in the CDW phase.}
\label{Fig:structure}
\end{figure*}

In recent years, the successful synthesis of two-dimensional (2D) van der Waals materials has led to novel ways of engineering flat bands\cite{Balents2020}. 
2D systems offer new degrees of freedom to study emergent behavior\cite{Geim2013,doi:10.1126/science.aac9439}.
Magic-angle twisted bilayer graphene is the paradigmatic example that exploits the weak interlayer van der Waals bonding to introduce a specific-angle rotation between the layers and artificially create a flat band\cite{PhysRevB.82.121407,bistritzer2011moire,cao2018correlated}. The electronic filling of this flat band can be controlled with an external electric gate between the layers, allowing the exploration of a rich phase diagram where unconventional superconducting states emerge at certain fillings\cite{cao2018unconventional,Oh2021}. 
Apart from that, the low dimensionality of van der Waals materials enhances the emergence of symmetry-broken orders such as charge density waves (CDW) that can be coupled to other phases\cite{doi:10.1021/acs.jpcc.9b08868,doi:10.1021/acs.jpcc.0c04913,doi:10.1021/acs.nanolett.2c04584}. Remarkably, CDW phases can induce reconstructions in the energy levels and create flat bands.
TaS$_2$ and TaSe$_2$ are the prototypical 2D compounds in which the formation of a CDW in the 1T phase (Fig. \ref{Fig:structure}a) leads to the emergence of a flat band at the Fermi level\cite{kratochvilova2017low,yu2017electronic,Chen2020}. 
It has been shown that this flat band is at half-filling and develops a magnetic insulating phase with spin $S=1/2$, both for TaS$_2$ and TaSe$_2$\cite{PhysRevLett.130.156401,doi:10.1021/acs.nanolett.3c02813}.
This phase has been considered as a promising candidate for the observation of a quantum spin liquid \cite{doi:10.1073/pnas.1706769114,PhysRevLett.121.046401,Ruan2021,Chen_2022}. It has also been used to engineer heavy fermions in artificial van der Waals heterostructures\cite{Vano2021}.
Moreover, superconducting states have been reported in the proximity of TaS$_2$ and TaSe$_2$. These have been induced by high-pressure\cite{Sipos2008}, chemical substitution of Ta\cite{PhysRevLett.109.176403,2022arXiv220305650W}, Li-ion intercalation between layers\cite{Yu2015}, and by ligand substitution in TaS$_{2-x}$Se$_x$\cite{liu2013superconductivity,PhysRevX.7.041054}. In these systems, the superconducting phase has been associated with the formation of a nearly commensurate CDW phase that drives the system to a metallic phase, allowing superconductivity to emerge\cite{Ang2015}. Moreover, superconductivity has been found experimentally in the vicinity of $x$=1.0 in the bulk TaSe$_{2-x}$Te$_x$ system\cite{PhysRevB.94.045131,doi:10.1073/pnas.1502460112}. However, the superconducting phase has not been fully characterized and the monolayer limit of this alloy remains unexplored.
 
In this work, we use a combination of \emph{ab initio} calculations and low-energy models to demonstrate that monolayer 1T-TaSe$_{2-x}$Te$_x$ undergoes a transition from a magnetic insulating phase to a non-magnetic metallic one. At intermediate concentrations $x\in[{0.846},{1.231}]$ the system becomes a magnetic metal in which the spin-polarized flat band is self-doped by the ligand substitution. We show that attractive interactions in this regime give rise to spin-triplet superconductivity. In particular, we observe that three different triplet-superconducting phases can occur as a function of the ligand substitution: nodal f-wave, topological chiral p-wave, and trivial chiral p-wave superconducting phases.

\section{Self-doped flat band}

\begin{figure*}
\centering
\includegraphics[width=\textwidth,draft=false]{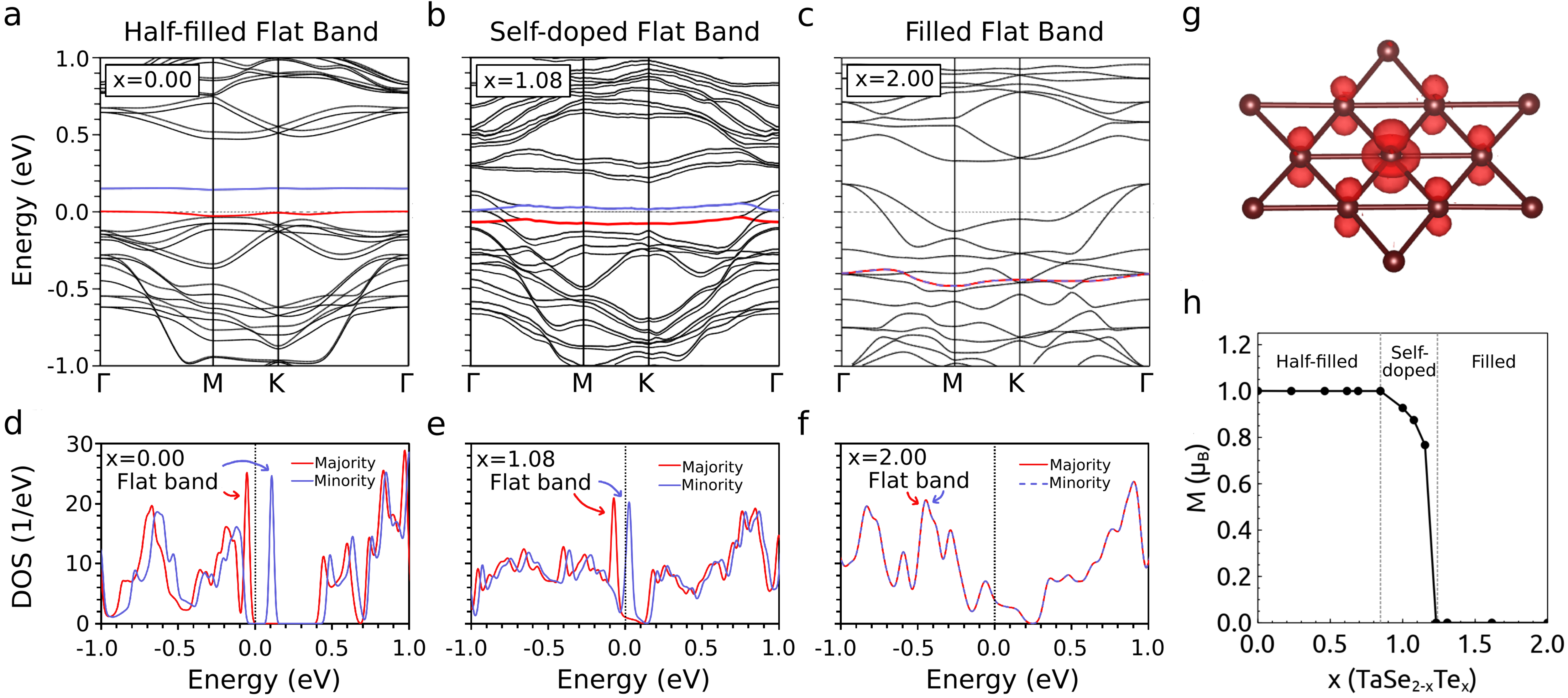}
\caption{Ab initio calculations in TaSe$_{2-x}$Te$_x$. Band structure (a)-(c) and DOS (c)-(f) of the majority and minority spin channels for different Te-doped cases: (a),(d) Half-filled flat band, (b),(e) doped flat band and (c),(f) filled flat band. The majority (minority) spin channel of the flat band is represented in red (blue). (g) Square of the wavefunction associated with the flat band at the $\Gamma$ point. (f) Magnetization phase diagram as a function of the amount of Te in the TaSe$_{2-x}$Te$_x$ system.}
\label{Fig:DFT}
\end{figure*}

We start our analysis performing density functional theory (DFT) calculations in TaSe$_{2-x}$Te$_x$ in the 1T $\sqrt{13}\times\sqrt{13}$ CDW phase \cite{kratochvilova2017low, BROUWER198051} also known as the star of David phase (see Fig. \ref{Fig:structure}a).\footnote{Details of the calculations can be seen in the Supplemental Material.} 
The symmetry of this CDW structure corresponds to space group no. 147, which is a subgroup of the normal state (NS) structure (space group symmetry no. 164). Three inequivalent Ta atoms are needed to describe this star-of-David CDW, represented in Fig. \ref{Fig:structure}a. There is a central Ta atom, six equivalent Ta atoms that form a hexagon surrounding it, and six equivalent Ta atoms at the points of the star of David. 
We refer to these as Ta1, Ta2, and Ta3 respectively (see Fig. \ref{Fig:structure}a). 
We have systematically analyzed TaSe$_{2-x}$Te$_x$ by modifying the Te-atom concentration $x$ by random-ligand substitution in the CDW unit cell (Fig. \ref{Fig:structure}a). 
The lattice parameters used in our calculations were the relaxed ones: 12.72 \text{\AA} for TaSe$_2$ and 13.29 \text{\AA} for TaTe$_2$. For the intermediate concentrations, the lattice parameters were linearly interpolated between these two limiting cases. For all concentrations, we optimized the atomic positions and observed that the equilibrium positions retain the star-of-David modulation. Moreover, we have performed phonon calculations in the CDW phase of TaSe$_2$, TaSe$_{0.92}$Te$_{1.08}$ and TaTe$_2$, which are representative cases of the three different electronic phases that we will explain later. All three calculations show stable phonon bands (Figs. \ref{Fig:structure}b, \ref{Fig:structure}c and \ref{Fig:structure}d respectively).
DFT calculations also show that the CDW phase is lower in energy than the NS for both the stoichiometric compounds, with an energy difference of $\sim$79 meV/Ta atom for TaSe$_2$ and $\sim$133 meV/Ta atom for TaTe$_2$, in agreement with recent analyses\cite{https://doi.org/10.1002/smll.202300262}. 
Experimentally, both monolayer TaSe$_2$ and TaTe$_2$ have been synthesized already. TaSe$_2$ in the monolayer limit has been shown to present the star-of-David pattern\cite{chen2020strong}, leading to all sorts of interesting physical properties governed by strong correlations\cite{zhang2020mottness}. TaTe$_2$ has also been grown as a monolayer, but it presents a much more complex pattern, with different CDW's present in the same nanosheet\cite{https://doi.org/10.1002/smll.202300262}. Moreover, a plethora of different structural phases has been reported as a function of $x$ in bulk  TaSe$_{2-x}$Te$_x$\cite{huisman1969polymorphism,vernes1998crystal,liu2013superconductivity}. Our \emph{ab initio} results indicate that ligand substitution by Te atoms maintains the dynamic stability of the star-of-David phase in monolayer 1T-TaSe$_{2-x}$Te$_x$ at all concentrations, which suggests that this phase could be synthesized under the right thermodynamic conditions.

We analyze now the evolution of the electronic structure as a function of the ligand substitution in TaSe$_{2-x}$Te$_x$. 
Figures \ref{Fig:DFT}a and \ref{Fig:DFT}d show the band structure for TaSe$_2$ and the corresponding density of states (DOS). We can observe that the CDW induces the formation of a flat band at half-filling that spin splits due to the dominant Coulomb interactions compared with the kinetic energy. 
The CDW unit cell presents a magnetization of 1 $\mu$$_B$, i.e. each star of David behaves as an individual spin $1/2$. 
From our DFT results, we see that the flat band has mainly Ta1- and Ta2-\textit{d}$_{z^2}$ orbital character, as can be seen in Fig. \ref{Fig:DFT}g where the square of the wavefunction associated to the flat band at the $\Gamma$ point is plotted. 
Our \emph{ab initio} calculations show that substituting Se by isoelectronic Te atoms induces electronic doping of the flat band and drives the system through three different phases. 
Representative band structures and DOS at different concentrations $x$ for the three phases are shown in Figs. \ref{Fig:DFT}(a,b,c) and \ref{Fig:DFT}(d,e,f) respectively. 
Moreover, these phases can be distinguished in Fig. \ref{Fig:DFT}h, where the evolution of the magnetization as a function of the ligand substitution $x$ is plotted. At low Te concentrations, i.e. 0$\leq x\leq$0.846 the flat band remains half-filled, and hence the system is in a magnetic insulating phase with a magnetic moment of 1 $\mu$$_B$ as for TaSe$_2$ (Figs. \ref{Fig:DFT}a and \ref{Fig:DFT}d). For intermediate concentrations, $0.846<x<1.231$, the minority spin channel of the flat band gets doped and the system enters into a magnetic metallic phase (Figs. \ref{Fig:DFT}b and \ref{Fig:DFT}e). Finally, for high Te concentrations, $1.231 \leq x\leq 2.0$, the flat band is fully filled and there is no splitting between the spin channels (Figs. \ref{Fig:DFT}c and \ref{Fig:DFT}f). In this regime, the system is in a non-magnetic metallic phase. Note that we have also traced the evolution of the flat band as a function of $x$ by analyzing the evolution of the wavefunction shown in Fig. \ref{Fig:DFT}g. We have observed that the character of this band does not get modified, which points out that ligand substitution does not cause hybridizations of the flat band with ligand states.
Therefore, the electronic doping of the flat band as a function of $x$ can be understood in terms of the electronegativity of the two ligands. 
Te is less electronegative than Se, thus providing less ionic character to the system. This causes the bands associated with the ligands to increase their energy and reach the Fermi level when increasing the Te concentration. This evolution can be seen in Figs. \ref{Fig:DFT}a,b,c. Since the flat band does not hybridize with these states, charge transfer is produced from the ligands to the flat band in a process of self-doping by isoelectronic ligand substitution. Let us recall that TaTe$_2$ and TaSe$_2$ are isoelectronic compounds, doping with Te does not add or remove electrons in the system, it only reorganizes the bands. 

\begin{figure}
\centering
\includegraphics[width=\columnwidth, draft=false]{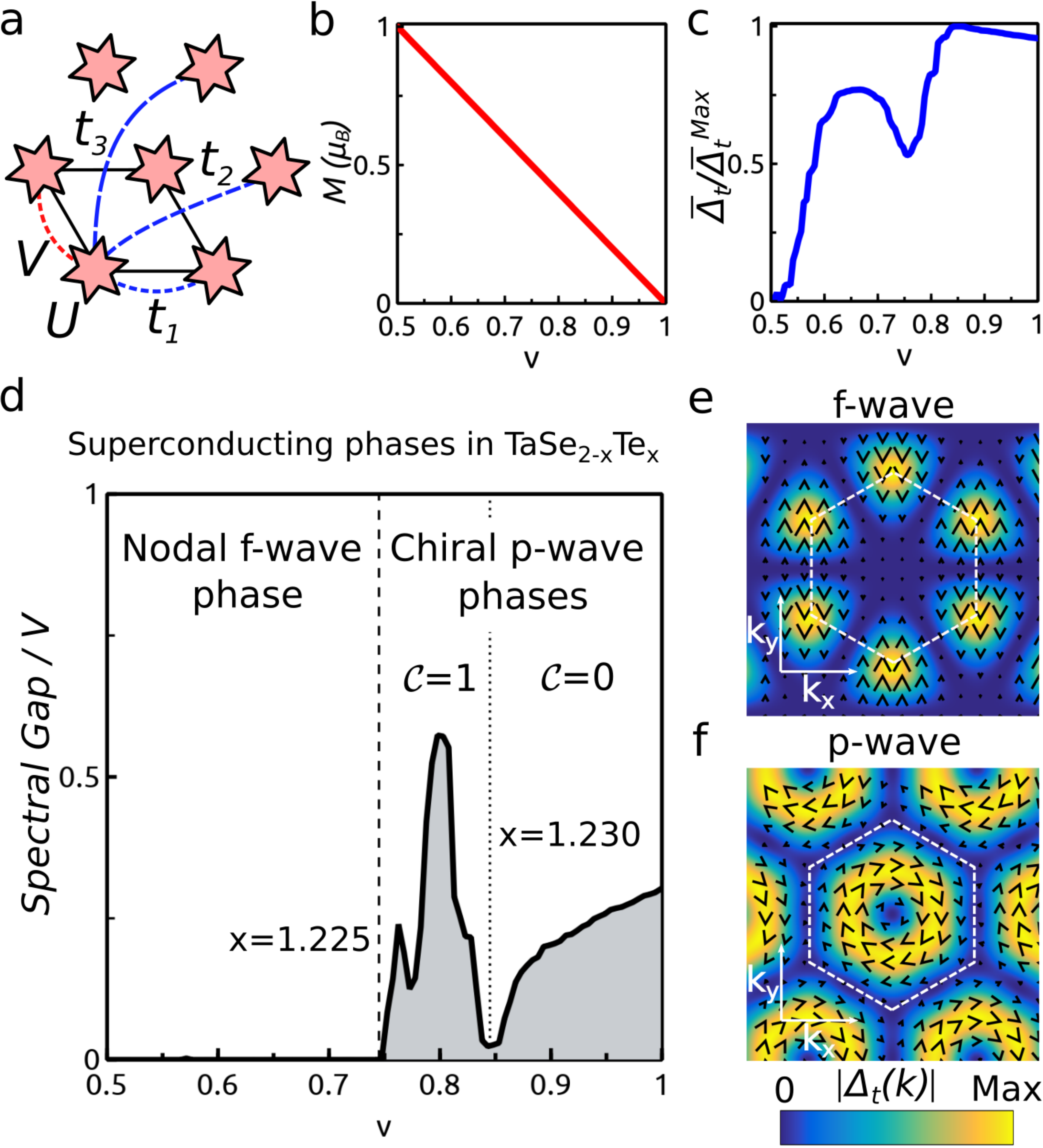}
\caption{(a) Scheme of the low-energy model used to study the self-doped flat band in TaSe$_{2-x}$Te$_x$. The orbital associated with the flat band is depicted with the red star forming a triangular lattice, $t_n$ correspond to hoppings to different neighbors, $U$ is the on-site repulsive Coulomb interaction, and $V$ is the attractive Coulomb interaction between first neighbors. (b) Magnetization as a function of electronic filling $\nu$. (c) Normalized spin-triplet superconducting order parameter $\bar \Delta_t$ as a function of $\nu$. (d) Normalized spectral gap as a function of $\nu$. (e,f) Spin-triplet superconducting order parameter in reciprocal space $|\Delta_t(\mathbf{k})|$ for the nodal f-wave (e) and the chiral p-wave (f) superconducting phases. The first Brillouin zone is depicted as the black-dashed hexagon. The real and imaginary parts of $\Delta_t(\mathbf{k})$ are plotted as a vector field for easy identification of the $f_{y(y^2-3x^2)}$ (e) and the $p_x + ip_y$ (f) symmetries of the superconducting phases.}
\label{Fig:model_phasediagram}
\end{figure}

\section{Spin-triplet superconductivity}

After having established from an \emph{ab initio} perspective that TaSe$_{2-x}$Te$_x$ undergoes a transition from a half-filled to a fully filled flat band, we analyze with a low-energy model derived from the DFT calculations the doped flat band phase. This regime is particularly interesting since it corresponds to a doped spin-polarized flat band in which Coulomb interactions dominate. From our first-principles calculations, we can extract a single-orbital Hamiltonian for the flat band in a triangular lattice:

\begin{equation}\label{eq:DFT_hamiltonian}
 \mathcal{H}_{0}=  \sum_{ s, i,j}t_{ij} c_{i,s}^{\dagger}c_{j,s}+U\sum_{i}c_{i\uparrow}^{\dagger}c_{i\uparrow}c_{i\downarrow}^{\dagger}c_{i\downarrow}, 
\end{equation}

where $c_{i,s}^{\dagger}$ ($c_{i,s}$) are the creation (annihilation) fermionic operator at site $i$ and for spin $s$, $t_{ij}$ are the hoppings between Wannier states of the flat band, and
$U$ is the effective on-site repulsive Coulomb interaction that leads to the spin splitting of the flat band. 
We can extract all the $t_{ij}$ hoppings and the $U$ from the DFT calculation\footnote{See the Supplemental material for further details} and the single orbital corresponds to that shown in Fig. \ref{Fig:DFT}g. 
Considering that the star of David in the family of Ta-based dichalcogenides is proximal to superconductivity\cite{Sipos2008,PhysRevLett.109.176403,2022arXiv220305650W,Yu2015,PhysRevX.7.041054,Ang2015}, bulk 1T-TaSe$_{2-x}$Te$_x$ displays a superconductive dome from $x=$ 0.8 to 1.3\cite{PhysRevB.94.045131,doi:10.1073/pnas.1502460112} and that ferromagnetic interactions (between stars) have been obtained for the parent compounds\cite{PhysRevB.105.L081106,jiang2021two}, this magnetic metallic phase in monolayer TaSe$_{2-x}$Te$_x$ unlocks the possibility for unconventional triplet superconductivity.
This occurs at the same regime where the magnetic metallic phase emerges in our DFT calculations for monolayer 1T-TaSe$_{2-x}$Te$_x$, and considering the presence of symmetry-breaking orders such as the CDW order or ferromagnetism, which are likely to induce a renormalization of the Coulomb interactions through electron-phonon coupling, charge, and spin fluctuations, we have added an effective attractive first-neighbor interaction $V$ to the Hamiltonian built from DFT (eq. (\ref{eq:DFT_hamiltonian})) 

\begin{equation}\label{eq:attractive_hamiltonian}
 \mathcal{H}=\mathcal{H}_{0}+V\sum_{\langle ij\rangle ss'}c_{i,s}^{\dagger}c_{i,s}c_{j,s'}^{\dagger}c_{j,s'}.
\end{equation}

A schematic of the low-energy model can be seen in Fig. \ref{Fig:model_phasediagram}a. This minimal model allows us to characterize the emergent superconductivity in monolayer 1T-TaSe$_{2-x}$Te$_x$ in case it survives from bulk \cite{PhysRevB.94.045131,doi:10.1073/pnas.1502460112}. The model can be solved as a function of the electronic filling of the flat band $\nu$ using a self-consistent mean-field approximation in the Nambu formalism including all the Wick contractions\cite{pyqula}. The correspondence between Te concentration $x$ in TaSe$_{2-x}$Te$_x$ and $\nu$ can be extracted from the DFT calculations\footnote{See the Supplemental material for further details}. 
We have obtained the ground-state solution for eq. (\ref{eq:attractive_hamiltonian}) as a function of $\nu$. We observe that two symmetry-broken orders get stabilized. First, a ferromagnetic order characterized by the magnetization $M=\langle c_{\uparrow}^{\dagger}c_{\uparrow}\rangle -\langle c_{\downarrow}^{\dagger}c_{\downarrow}\rangle$ as the order parameter. This is a consequence of the repulsive on-site Coulomb interaction $U$. Increasing the filling of the flat band decreases the magnetization (Fig. \ref{Fig:model_phasediagram}b) as obtained in the DFT calculation. 
Secondly, a spin-triplet superconducting phase emerges due to the attractive interactions $V$ coexisting with a magnetic state. We can observe that the order parameter for spin-triplet superconductivity
$\bar \Delta_t = \sqrt{\int_{BZ} |\Delta_t (\mathbf k)|^2 d^2 \mathbf k}$, where
$\Delta_t (\mathbf k)= \langle c_{k,\uparrow}^{\dagger}c_{-k,\uparrow}^{\dagger}\rangle$,  is finite for every $\nu$ in the doped flat band regime (Fig. \ref{Fig:model_phasediagram}c).
The spin-triplet phase is a direct consequence of the spin-polarized flat band, which by symmetry only allows triplet phases to occur. 
Spin-triplet superconductors are scarce in nature since the coupling occurs between electrons with aligned spins, and they are more resilient to magnetic fields. Traditionally, this kind of superconductivity is found in rare-earth compounds with localized f electrons in the context of heavy-fermion systems\cite{Saxena2000,Aoki2001,doi:10.1126/science.aav8645,doi:10.1126/science.1187943,doi:10.1126/science.1248552}. We show here that monolayer 1T-TaSe$_{2-x}$Te$_x$ could establish a different family of spin-triplet superconductors. 

We can analyze the spectral gap associated with the spin-triplet superconducting state (Fig. \ref{Fig:model_phasediagram}d). This allows us to distinguish three different spin-triplet phases as a function of the flat band electronic filling. For Te concentrations of 0.846$\leq x\leq$1.225,  the spin-triplet phase is gapless, i.e. the superconducting gap opens only in part of the Fermi surface. We can observe the dependence in reciprocal space of the spin-triplet order parameter $|\Delta_t(\mathbf{k})|$ for this phase in Fig. \ref{Fig:model_phasediagram}e. We can see that the superconducting gap opens around the $K$ high-symmetry points, but not along the $\Gamma-M$ path\footnote{The evolution of the Fermi surface when including attractive interactions and the band structure for these gapless phases are included in the supplemental material.}. This symmetry corresponds to a $f_{y(y^2-3x^2)}$ nodal f-wave superconductor, which has been reported in TaS$_2$ for the 1H phase\cite{2021arXiv211207316V} and as a
competing phase in NbSe$_2$\cite{Wan2022}. For Te concentrations of 1.225$\leq x\leq$1.231,  the spin-triplet superconductor becomes gapped (Fig. \ref{Fig:model_phasediagram}d). The symmetry in this case corresponds to a chiral p-wave $p_x + ip_y$ spin-triplet phase (Fig. \ref{Fig:model_phasediagram}f). 
These nodal f-wave and the fully gapped p-wave phases could be experimentally distinguished based on the V or U-shaped gap in their spectra \cite{Kim2022, 2021arXiv211207316V} \footnote{See the Supplemental Material for a calculation of the spectrum of each spin-triplet phase.}

\begin{figure}[ht]
\centering
\includegraphics[width=\columnwidth, draft=false]{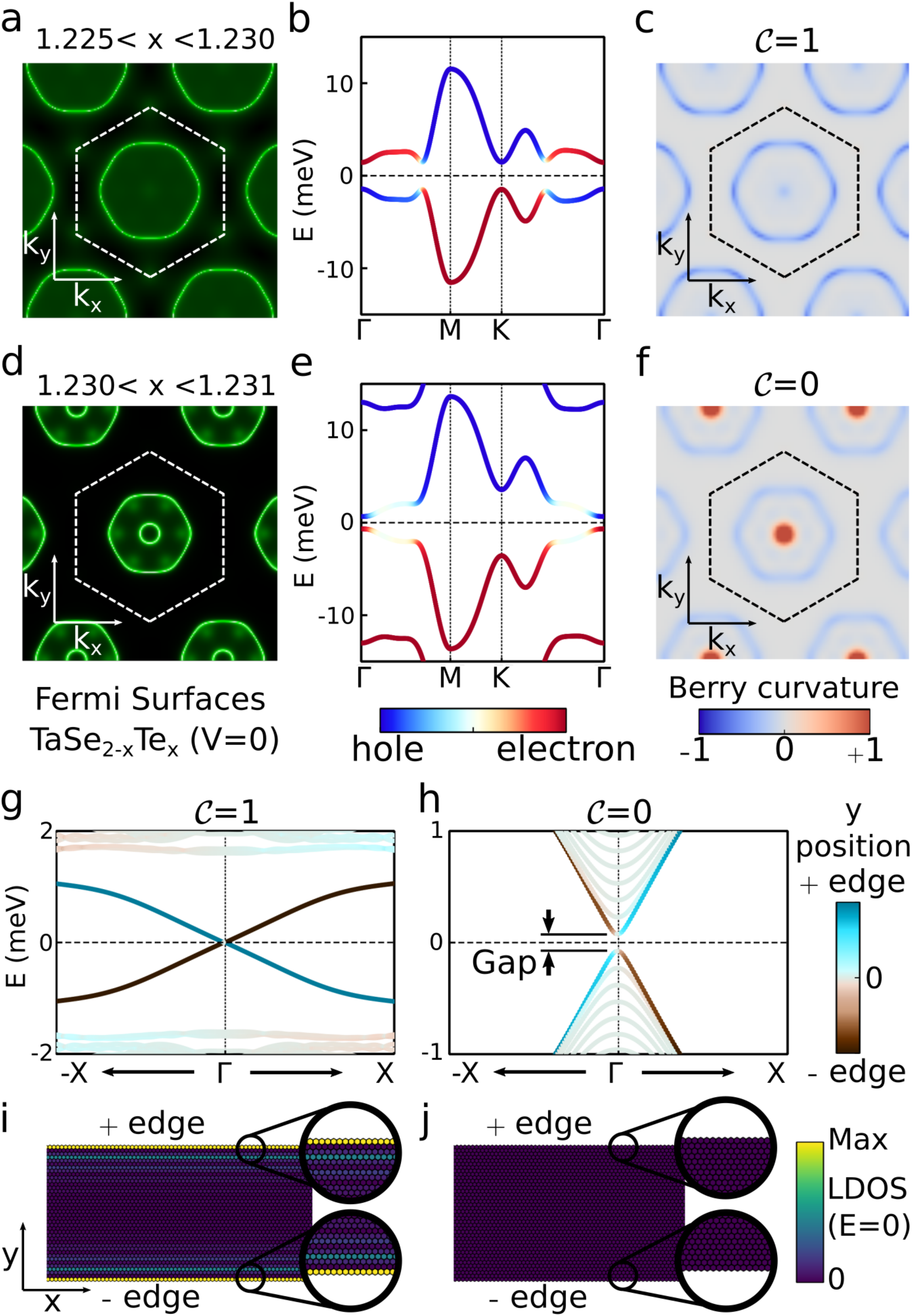}
\caption{Fermi surface in the absence of attractive interactions ($V=$0) at different concentrations for the topologically non-trivial (a) and trivial (d) chiral p-wave phases. Band structure (b,e) and corresponding Berry curvature (c,f) for the topologically non-trivial and trivial chiral p-wave phases for $V<0$. Band structure (g,h) and corresponding Local Density of States at zero energy LDOS($E=0$) (i,j) for a ribbon (infinite in the x direction and finite in the y direction) in the topological and trivial chiral p-wave phases respectively.}
\label{Fig:model_gappedSC}
\end{figure}

In the fully-gapped p-wave regime, a Chern number $\mathcal{C}$ can be defined, and the topology of the spin-triplet superconductor be analyzed (Fig. \ref{Fig:model_phasediagram}d). For Te concentrations of 1.225$\leq x\leq$1.230, $\mathcal{C}=1 $ and the system is in a topological spin-triplet superconducting state. The cusp of the superconducting dome in this topological phase occurs at $x=1.229$. For Te concentrations of 1.230$\leq x\leq$1.231, $\mathcal{C}=0 $ and the p-wave phase is trivial. 
The topological transition in the chiral p-wave phase stems from a Lifshitz transition as shown in Fig. \ref{Fig:model_gappedSC}. For the topological phase, the Fermi surface  in the absence of interactions (Fig. \ref{Fig:model_gappedSC}a) shows a pocket around the $\Gamma$ point. When attractive interactions are included, a superconducting gap opens in the band structure (Fig. \ref{Fig:model_gappedSC}b) and the Berry curvature has the same sign for every k-point leading to a finite Chern number $\mathcal{C}=1 $ (Fig. \ref{Fig:model_gappedSC}c). 
In the trivial phase, a second pocket around the $\Gamma$ point emerges in the Fermi surface in the absence of interactions due to the filling of the flat band (Fig. \ref{Fig:model_gappedSC}d). When attractive interactions are included, a superconducting gap opens in the band structure (Fig. \ref{Fig:model_gappedSC}e) and, as a result of the new pocket, a second band inversion occurs at the $\Gamma$ point.  This introduces a source of opposite-sign Berry curvature at the $\Gamma$ point leading to a net zero Chern number $\mathcal{C}=0 $ (Fig. \ref{Fig:model_gappedSC}f). 
We demonstrate the bulk-boundary correspondence of the two phases by analyzing a ribbon (infinite in the $x$-direction and finite in the $y$-direction) in each regime. In the topological phase, the ribbon shows counter-propagating zero-energy modes as shown in the band structure in Fig. \ref{Fig:model_gappedSC}g. Each mode is localized in a different edge, as shown by the $y$-position operator in the band structure and the Local Density of States at zero energy (LDOS(E=0)) in Fig. \ref{Fig:model_gappedSC}i. In the trivial phase, the ribbon is gapped (Fig. \ref{Fig:model_gappedSC}h) and there are no zero-energy modes (Fig. \ref{Fig:model_gappedSC}j). The edge modes associated with the topological phase could be experimentally detected with a Scanning Tunneling Microscope \cite{Kezilebieke2020}.

\section{Conclusions}
In conclusion, we have demonstrated that TaSe$_{2-x}$Te$_x$ undergoes a transition from a magnetic insulating phase to a non-magnetic metallic one as a function of $x$. For $x\in[{0.846},{1.231}]$ the system becomes a magnetic metal in which a spin-polarized flat band is self-doped by the isoelectronic ligand substitution. In this regime, we have shown that attractive interactions promote three different spin-triplet superconducting phases: a nodal f-wave, a chiral topological p-wave, and a chiral trivial p-wave phase. These can be accessed by ligand-substitution engineering at particular concentrations. Our results demonstrate that monolayer TaSe$_{2-x}$Te$_{x}$ is a potential platform for a variety of unconventional spin-triplet superconducting states and strongly correlated driven flat band phenomena.

\section*{Data availability statement}

All data that support the findings of this study are included within the article (and any supplementary files).

\section*{Acknowledgements}
This work was supported by the MINECO of Spain project PID2021-122609NB-C22, the Academy of Finland Projects No. 331342, No. 336243 and No. 349696 and the Jane and Aatos Erkko Foundation. J. P. thanks MECD for the financial support received through the ``Ayudas para contratos predoctorales para la formación de doctores" grant PRE2019-087338. 
We acknowledge the computational resources provided by the Galician Supercomputing Center (CESGA) and the Aalto Science-IT project.

\bibliography{dstar}

\end{document}


\title{Supplemental Material: Self-doped flat band and spin-triplet superconductivity in monolayer 1T-TaSe$_{2-x}$Te$_{x}$}

\author{Jan Phillips}
  \email{j.phillips@usc.es}
\affiliation{Departamento de F\'{i}sica Aplicada,
  Universidade de Santiago de Compostela, E-15782 Campus Sur s/n,
  Santiago de Compostela, Spain}
\affiliation{Instituto de Materiais iMATUS,
  Universidade de Santiago de Compostela, E-15782 Campus Sur s/n,
  Santiago de Compostela, Spain}   
\author{Jose L. Lado}
\affiliation{Department of Applied Physics, Aalto University, 02150 Espoo, Finland}
\author{Víctor Pardo}
\affiliation{Departamento de F\'{i}sica Aplicada,
  Universidade de Santiago de Compostela, E-15782 Campus Sur s/n,
  Santiago de Compostela, Spain}
\affiliation{Instituto de Materiais iMATUS,
  Universidade de Santiago de Compostela, E-15782 Campus Sur s/n,
  Santiago de Compostela, Spain}  
\author{Adolfo O. Fumega}
\email{adolfo.oterofumega@aalto.fi}
\affiliation{Department of Applied Physics, Aalto University, 02150 Espoo, Finland}

\maketitle

\section{Computational details in the DFT calculations}

We have performed ab initio electronic structure calculations based on Density Functional Theory (DFT) using an all electron full-potential code (WIEN2k\cite{WIEN2k}). The generalized gradient approximation (GGA)\cite{perdew1996generalized} was used as the exchange-correlation term for all of our calculations. A converged \textit{k}-mesh and a value of R$_{mt}$K$_{max}$=7.0 was used for the relaxations of the limit cases TaSe$_2$ and TaTe$_2$, with the values used for the muffin-tin radii being R$_{MT}$(Ta)=2.45, R$_{MT}$(Se)=2.17 and R$_{MT}$(Te)=2.34.
The harmonic phonon spectrum of the charge-density wave (CDW) phase of TaSe$_2$ and TaTe$_2$ was computed using the real-space supercell approach\cite{phonopy}. Taking the CDW structure as the unit cell, we have performed calculations of a $2\times2$ supercell at the $\Gamma$ point ($1\times1\times1$ \textit{k}-mesh). We have used the VASP \cite{kresse1993vasp,kresse1996efficiency,kresse1996efficient}code for this task. We set the cutoff energy for the plane-wave-basis set (ENCUT) to 300 eV for both compounds and the cut-off energy of the plane wave representation of the augmentation charges (ENAUG) to 450 eV for TaSe$_2$ and 400 eV for TaTe$_2$.

\begin{figure}[h]
\begin{center}
\includegraphics[width=0.8\columnwidth, draft=false]{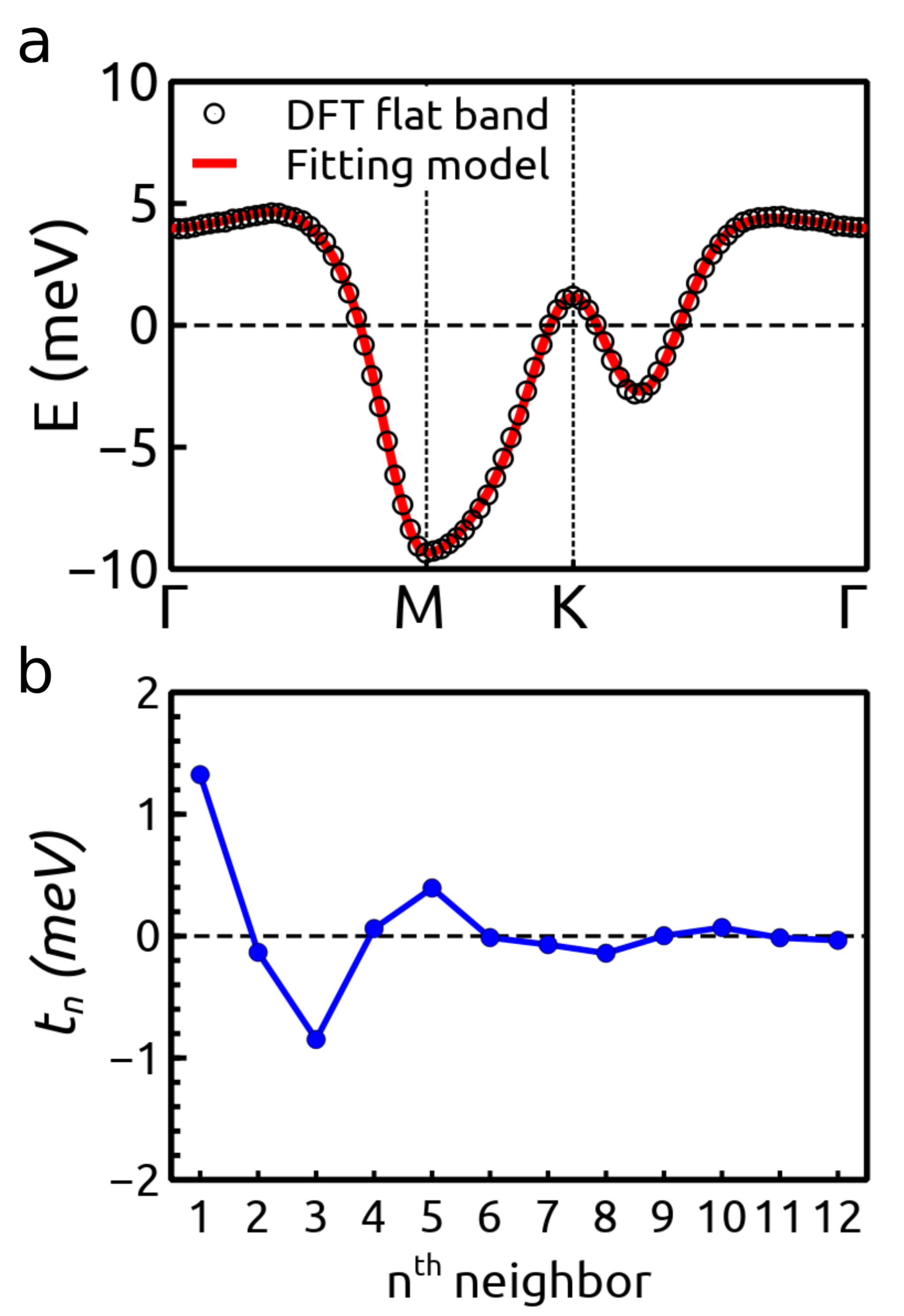}
\caption{(a) Fitting of the flat band obtained from DFT to eq. (1). (b) Hopping parameters $t_n$ for different neighboring distances.}
\label{Fig:dft_fit}
\end{center}
\end{figure}

\section{Fitting of the low energy model to the DFT flat band}

In this section, we show the fitting of the flat band obtained from DFT to eq. (1) in the main text. The fitting is shown in Fig. \ref{Fig:dft_fit}a. The hopping parameters $t_n$ for different neighboring distances are plotted in Fig. \ref{Fig:dft_fit}b. We have also obtained a $U=149.6$ meV that leads to the spin splitting obtained from the DFT calculations.

\begin{figure}[h]
\begin{center}
\includegraphics[width=0.8\columnwidth, draft=false]{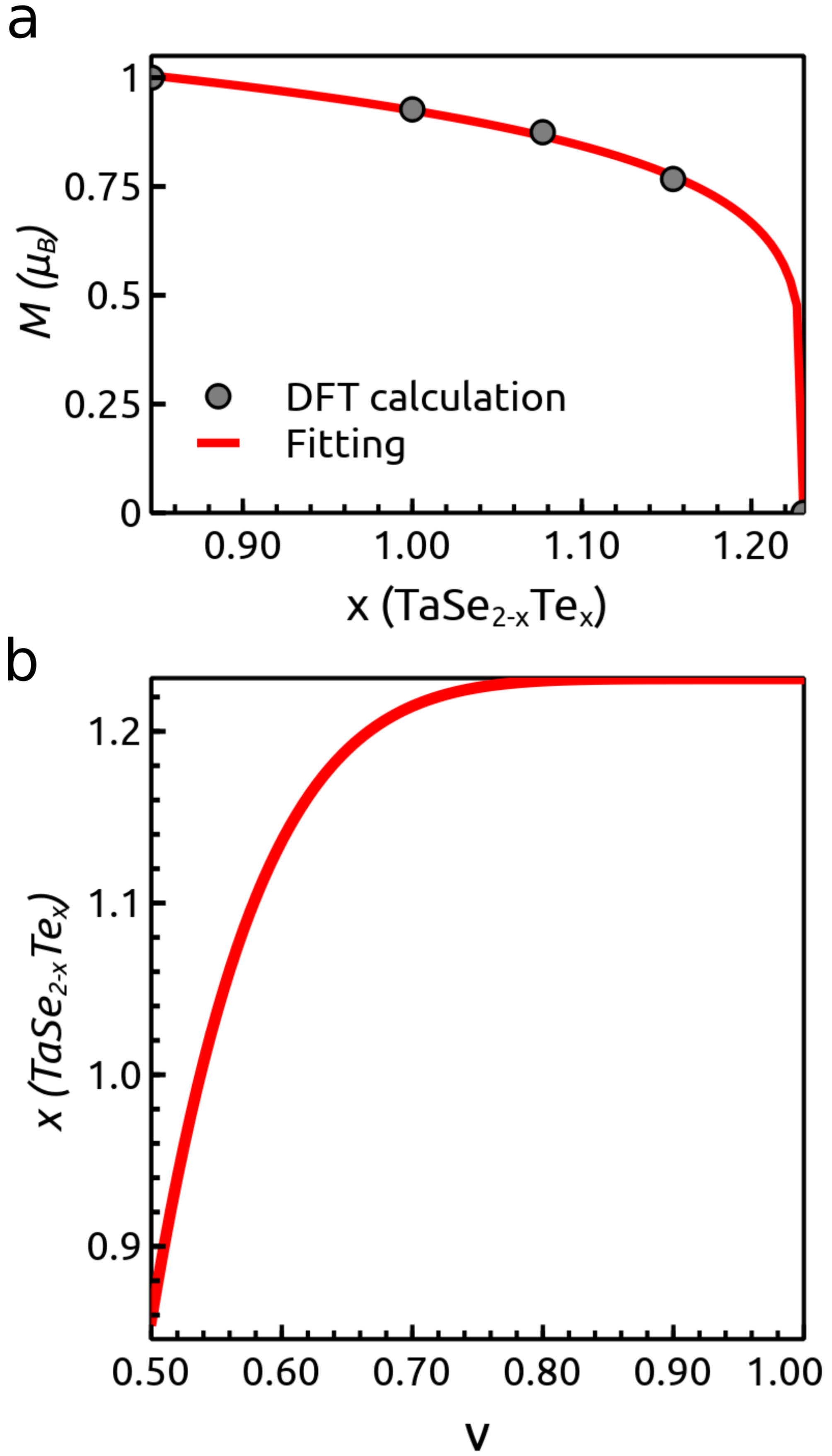}
\caption{(a) Fitting of eq. (\ref{eq:M_x}) to the DFT calculations. (b) relationship between  the electronic filling of the flat band $\nu$ and the content of Te $x$ in TaSe$_{2-x}$Te$_{x}$. }
\label{Fig:x_nu}
\end{center}
\end{figure}

\begin{figure}[h]
\begin{center}
\includegraphics[width=0.8\columnwidth, draft=false]{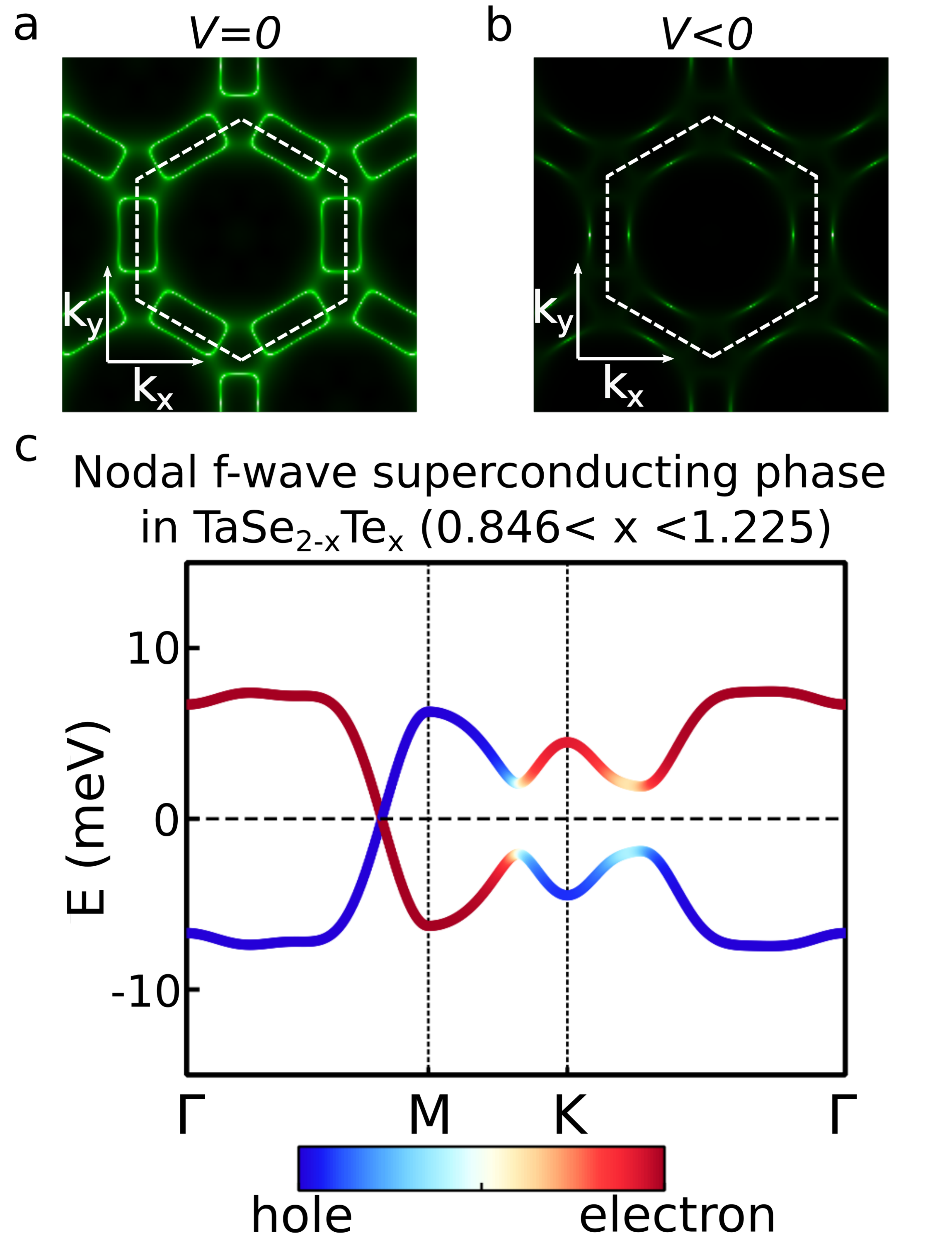}
\caption{Fermi surface for the nodal f-wave regime, 0.846$\leq x\leq$1.225, in the absence of attractive interactions ($V=0$) (a) and when attractive interactions are included ($V<0$) (b). (c) Representative band structure of the nodal f-wave phase. The bands are colored according to the electron operator.}
\label{Fig:gapless_phase}
\end{center}
\end{figure}

\section{Correspondence between ligand substitution and electronic filling}

The correspondence between the filling $\nu$ of the flat band studied in the low-energy model (eq. (2) in the main text) and the ligand substitution $x$ that results from the DFT calculations can be traced. From the different stoichiometries computed in DFT, we have obtained the total magnetization $M$ as a function of $x$.  
At low Te concentrations, i.e. 0$\leq x\leq$0.846 the flat band is half-filled $vu=0.5$, and $M=$1 $\mu$$_B$. For high Te concentrations, 1.231$\leq x\leq$2.0, the flat band is filled $vu=1.0$, and $M=$0 $\mu$$_B$. The self-doped flat band regime occurs for intermediate concentrations, 0.846$<x<$1.231. DFT calculations show that the total magnetization decays from 1 to 0 $\mu$$_B$  as a function of $x$. We can fit the DFT points to the following equation that gives the transition for this kind of transition:

\begin{equation}\label{eq:M_x}
    M=\alpha(1.231-x)^{\delta},
\end{equation}

where $\alpha=1.173$ and $\delta=0.162$ are the parameters that we obtain from the fitting to the DFT results (Fig. \ref{Fig:x_nu}a). From eq. (\ref{eq:M_x}) and considering that $M$ decreases linearly from 1 to 0 $\mu$$_B$  as a function of $\nu$ for $\nu\in[{0.5},{1.0}]$ (Fig. 2b in the main text). We can obtain the relationship between $\nu$ and $x$ shown in Fig. \ref{Fig:x_nu}b
 
\begin{equation}\label{eq:x_nu}
    x=1.231-\left(\frac{2(1.0-\nu)}{\alpha}\right)^{1/\delta}.
\end{equation}

\section{Nodal f-wave phase}

In this section, we include supplemental material for the nodal f-wave superconducting phase that occurs for Te concentrations in the range 0.846$\leq x\leq$1.225. 
Figure \ref{Fig:gapless_phase}a shows the Fermi surface in this doping regime in the absence of attractive interactions ($V=0$). We can see that rectangular-like pockets are formed around the M points. When attractive interactions are included a superconducting gap opens around the K high-symmetry points, but not along the $\Gamma$-M path as can be seen in Figs. \ref{Fig:gapless_phase}b and \ref{Fig:gapless_phase}c. This leads to the nodal f-wave phase.

\section{Density of States of the nodal and the fully gapped superconducting orders}

\begin{figure}[h!]
\centering
\includegraphics[width=\columnwidth, draft=false]{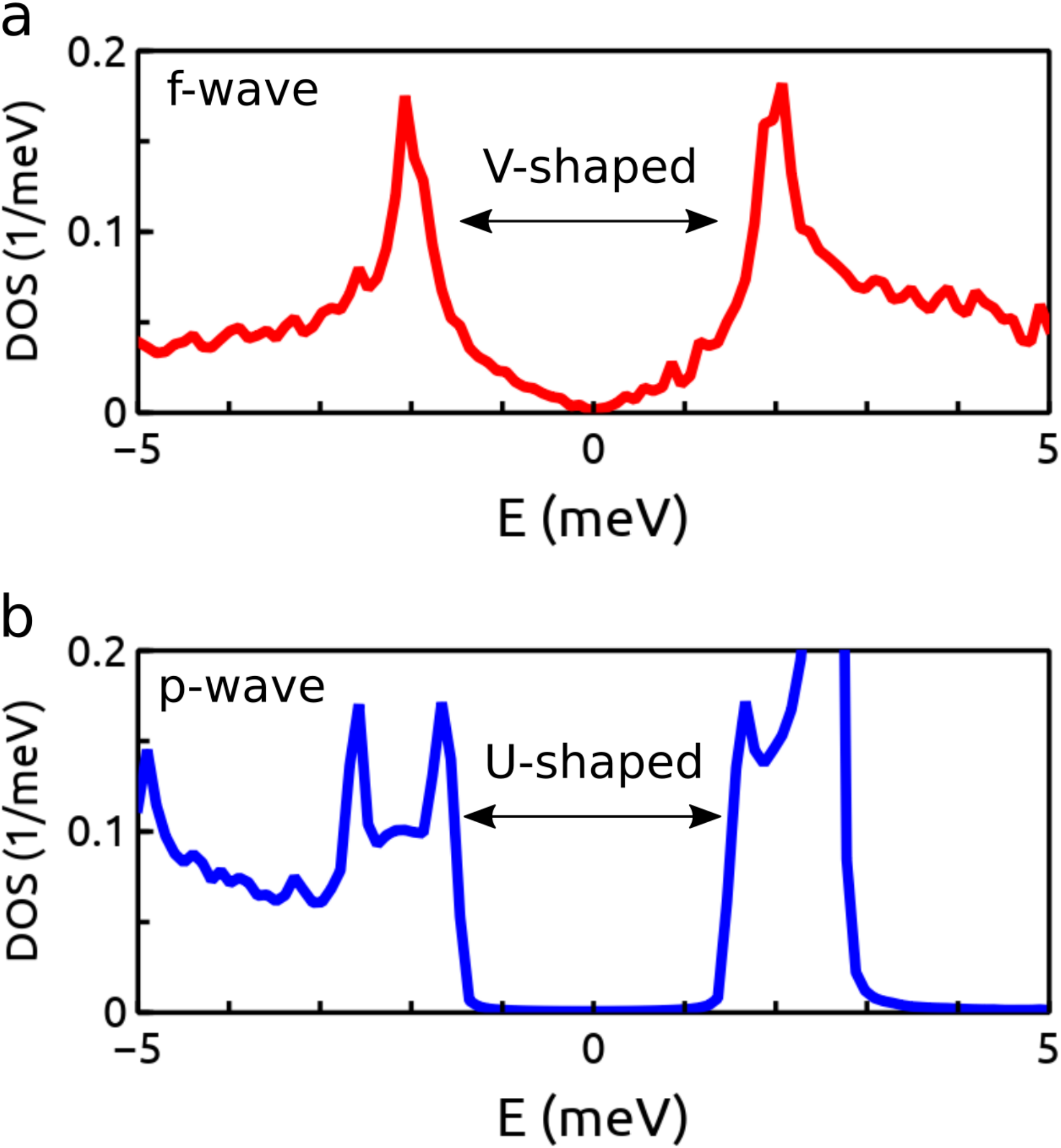}
\caption{Density of states (DOS) showing a V-shaped spectrum in the case of nodal f-wave phase (a) and a U-shaped gap in the case of the fully gapped p-wave phase (b).}
\label{Fig:VUshapeSCgap}
\end{figure}

In this section, we provide the density of States (DOS) that allows us to analyze the characteristic features of the spectra for the spin-triplet superconducting phases (Fig. \ref{Fig:VUshapeSCgap}). We can observe that the nodal f-wave phase develops a V-shaped spectrum (Fig. \ref{Fig:VUshapeSCgap}a) while the fully gapped p-wave phase develops a U-shaped gap (Fig. \ref{Fig:VUshapeSCgap}b). This analysis could be performed with a scanning tunneling microscope \cite{Kim2022, 2021arXiv211207316V}, allowing us to distinguish between these two superconducting phases based on the V or U shape of their gap. 

\bibliography{dstar}